# On Gaugino Dominated Dark Matter

**S.E. Ennadifi[1*], E.H. Saidi[1,2+]**

[1]*Laboratory of High Energy Physics, Modeling and Simulation, Faculty of Sciences, Rabat, Morocco*
[2]*Centre of Physics and Mathematics CPM-CNESTEN, Rabat, Morocco*

## Abstract

Using the neutral gauginos of $SU_L(1) \times U_Y(1)$ and hybridization ideas below the GUT scale, we approach the Dark Matter particle within the Minimal Supersymmetric Standard Model. In the energy range $M_{GUT} - M_Z$ where supergravity effects can be ignored, it is proposed that such DM particle could be interpreted in terms of a mixture of Bino and Wino states with a lower bound mass $M_{DM} \geq 65GeV$ not far above the electroweak scale to account for the observed Dark Matter density. We establish the theoretical origin of this particle and study as well its compositeness and its mass bound.

........................................................

*ennadifis@gmail.com
+h-saidi@fsr.ac.ma

# 1. Introduction

Precision cosmology data identify Dark Matter (DM) as the main building block for all structures in the Universe [1,2,3]; however they do not discriminate among the several candidates discussed in the literature namely MACHO's [4], Axions [5], gluinos [6,7], the lightest sneutrino [8,9], which some of them have been excluded experimentally [10,11], and weakly interacting massive particles (WIMP's) [12]. Although alternative explanations in terms of modified gravity (MOND) [13,14] cannot be ignored, they can hardly be reconciled with the most recent astrophysical observations [15,16] without requiring additional matter beyond the observed baryons [17].

Solving this puzzle is one of the greatest challenges in modern physics. This problem, long regarded in astrophysics and cosmology, is now deeply rooted in high energy physics since it is in the context of theories beyond the Standard Model (SM) of elementary particles [18] where the brightest candidates for DM arise. New weakly interacting massive particles [19] are well motivated by particle physics theory opening the door for particle physics beyond the SM. The Minimal Supersymmetric Standard Model (MSSM), the best motivated extension of the SM, provides a good candidate for the DM component of the Universe in terms of Lightest Supersymmetric Particle (LSP) if R-parity $R_P$ is conserved [20]. In simple realistic supersymmetric models, where supersymmetry is broken at the weak scale all SM particles must have superpartners with masses $\leq 1TeV$. These sparticles, existed in the early universe in thermal equilibrium with the ordinary SM particles. As the universe cooled and expanded, the heavier sparticles could no longer be produced, and they eventually annihilated or decayed into LSP's. Some of the LSP's pair-annihilated into final states not containing sparticles.

Given the outstanding advances in DM detection experiments, as well as the forthcoming onset of the LHC experiments, an exciting near future can be anticipated in which this enigma might start being unveiled. In this paper, we develop an approach to the DM weakly interacting massive particle by using the pure gaugino sector of MSSM and a hypothesis on hybridization of the wave functions of the abelian gauginos of the MSSM gauge group. Guided by LHC experiments, we will mainly focus on the $SU_C(3) \times SU_L(1) \times U_Y(1)$ Yang-Mills sector of MSSM although a full picture should include supergravity since, along with gauginos, it is a potential candidate for DM through the gravitino sector. To proceed, we first recall briefly the main evidence for DM. Then, restricting to the gaugino sector of MSSM and using wave function language, we present our proposal for DM particle in terms of hybridization of the Bino $\breve{B}_\alpha$ and the Wino $\widetilde{W}_\alpha^{\circ}$ fields, the supersymmetric partners of the $B_\alpha$ and $W_\alpha$ gauge particles mediating the $U_I(1) \times U_Y(1)$ gauge interaction sector of the SM. Next, we focus on the identification of the DM particle and its basic properties by varying the scale energy ε from the grand unification scale $M_{GUT}$ to the electroweak $M_Z$ one, then we deduce the lower bound of DM particle mass $M_{DM}$ by translation into the electroweak scale. We end this study by some comments and concluding remarks.

# 2. The problematic

Historically the observational evidence for the existence of DM came only from galactic dynamics. Presence of DM can be inferred from gravitational effects on visible matter. According to present observations of universe structures larger than galaxies as well as Big Bang cosmology inspired models, DM accounts for the vast majority of mass in the observable universe [21],

$$Universe \equiv DE)_{\sim 70\%} + DM)_{\sim 25\%} + SM)_{\sim 5\%} \tag{2,1}$$

where DM stands for non baryonic dark matter and DE for dark energy. Although the microscopic composition of DM remains a mystery, it is clear that it cannot consist of any elementary particles which have been discovered so far. Since this exotic matter strongly exists in the form of non-baryonic matter, the MSSM is therefore expected to provide a good candidate for the DM component of the Universe, thanks to the $R_P$ conservation. The current largely adopted view on DM implies that it is mostly made of WIMP's [12]. Although these are hypothetical massive particles that scarcely interact with regular matter, making them very hard to detect, a huge effort has been made to discover them on a global scale. DM component must be then an undiscovered, massive, stable or long lived particle. So we envisage that is, as it is almost widely accepted, supersymmetric, light, electrically neutral ($Q_{em} = 0$) and non-colored ($C = 0$). This limits our work to the following abelian dark sector

$$G_{Dark} \sim U^{\otimes n}(1)_{C=Q_{em}=0}. \qquad (2,2)$$

If weak-scale supersymmetry is realized in nature, the LSP plays a special role in the search for supersymmetry at colliders. All heavier particles rapidly cascade decay to the LSP, since this occurs in all supersymmetric events, the nature of the LSP and its behavior are decisive for all supersymmetric signatures at the LHC. In the MSSM, the most widely studied candidate the lightest neutralino $\widetilde{N}$, which is a very promising DM candidate [22] in large region of the parameter space, and, if one of its mass parameter contents is much lighter than the others, the LSP will be predominately of this form.

For the remainder of this paper, we will restrict ourselves to the case of gaugino dominated LSP, especially a typical Bino-like content $\Omega_{\breve{B}}$ as it often emerges from minimal supergravity boundary conditions on the soft parameters which tend to require it in order to get correct electroweak symmetry breaking. Although Bino DM generally gives higher DM density, it could be decreased by considering just enough Wino $\Omega_{\widetilde{W}^\circ}$ content allowing for natural cold DM in accordance with astrophysical datas.

Below, in the MSSM framework, we develop an approach to DM through the idea of the evolution of the abelian gaugino hybrids state with the scale energy down to the electroweak scale where the Higgsino sector contributes $\Omega_{\widetilde{H}^\circ}$ to DM state. Using specific properties where gauge quantum numbers captured by matter in adjoint representations take trivial values makes the detection of their interactions a complicated task and where LHC experiments are expected to bring more insight.

## 3. Dark Matter building

Restriction to the neutral gauginos sector of MSSM as the first source of DM makes the derivation of DM particle more tractable,

$$G_{Dark} = U_I(1) \times U_Y(1). \qquad (3,1)$$

In fact, besides their masses generated by supersymmetry breaking, the abelian gauginos $\widetilde{\Psi}_\alpha = \left(\breve{\Psi}_a, \breve{\Psi}_{\dot{a}}\right)$ (4D Majorana fermions) have the same quantum numbers under $SU_C(3) \times U_I(1) \times U_Y(1)$; they belong to uncharged (real) representations of the gauge symmetry. Restricting to the abelian subgroup (3,1), we have:

$$Bino: \breve{B}_\alpha \sim (1,1)_0,$$
$$Wino: \widetilde{W}^\circ \sim (1,1)_0 . \qquad (3,2)$$

In absence of supergravity, the Bino $\breve{B}_\alpha$ of the $U_Y(1)$ sub-invariance and the Wino $\widetilde{W}^\circ$ of the symmetry $U_I(1) \times U_Y(1)$, and eventually the two gluinos in the $U^2(1)$ Cartan sector of the $SU_C(3)$ color symmetry, are good candidates to dominate the DM brick. Focusing on these two massive particles in the $G_{Dark}$ gauge sector of MSSM, we have to deal with the two particle states,

$$|\tilde{B}\rangle_{C=Q_{em}=0}, \ |\widetilde{W}^\circ\rangle_{C=Q_{em}=0}. \qquad (3,3)$$

In this description, the field particles $\tilde{B}$ and $\widetilde{W}^\circ$ are Majorana spinors standing respectively for the wave functions associated with the gauginos $\left(\breve{B}_a, \breve{B}_{\dot{a}}\right)$ and $\left(\widetilde{W}^\circ_a, \widetilde{W}^\circ_{\dot{a}}\right)$. In addition to the fact that experimentally are hardly distinguishable, the trivial quantum numbers of the $|\breve{B}\rangle$ and $|\widetilde{W}\rangle$ states let understand that gauge invariance and supersymmetry do not prevent the existence of mixed states of these "*twin sparticles*" in the linear combination form,

$$|\widetilde{N}_+\rangle = \mathrm{f}|\breve{B}\rangle + \mathrm{g}|\widetilde{W}^\circ\rangle,$$
$$|\widetilde{N}_-\rangle = -\mathrm{f}|\breve{B}\rangle + \mathrm{g}|\widetilde{W}^\circ\rangle \qquad (3,4)$$

even before the electroweak symmetry breaking $SU_L(1) \times U_Y(1) \rightarrow U_{em}(1)$ where the two Higgsinos compete after getting the appropriate gauge quantum numbers

$$|\widetilde{H}^{\circ}\rangle_{C=Q_{em}=0}. \qquad (3,5)$$

Since $\widetilde{N}_{\pm}$ are still eigenstates of the $G_{Dark}$ gauge invariance and that Majorana condition still holds provided that the coefficients f and g are real which moreover obey the normalization condition

$$\langle \widetilde{N}_{\pm} | \widetilde{N}_{\pm} \rangle = \mathrm{f}^2 + \mathrm{g}^2 = 1,$$

$$\langle \widetilde{N}_{\pm} | \widetilde{N}_{\pm} \rangle = 0. \qquad (3,6)$$

On the other hand seen that the hybridization property (3,4) as well as the normalization condition (3,6) should hold for the energy band where MSSM is supposed to govern the dynamics of the universe evolution, we will assume moreover that the real coefficients f and g depend somehow on the scale energy $\varepsilon$, that is:

$$\mathrm{f} = \mathrm{f}(\varepsilon), \; \mathrm{g} {=} (\varepsilon) \qquad (3,7)$$

with $\varepsilon$ belonging more a less to the range $M_{GUT} - M_Z$ and where it plays a similar role as in the renormalization group equation for gauge coupling constants. In this picture, the DM component we are looking for depends on the scale energy and other extra moduli fixed by dimensional arguments; one of them, denoted τ, will be inferred in a moment; see eq (3.10). Within this view, let us now focus on the explicit building of the states (3,4) in terms of the scale energy. A priori, there are infinitely many solutions for the normalization condition (3,6) since the 2×2 rotation matrix is orthogonal, having a family of one parameter solution. But here we need to solve the constraint relation (3,6) in an explicitly energy dependent manner that could somehow describe a quasi-realistic model. Therefore seen that the involved coefficients solve the orthogonality condition of the $SO(2,R)$ rotation symmetry in the space generated by their associated wave functions,

$$R = \begin{pmatrix} \mathrm{f} & \mathrm{g} \\ -\mathrm{g} & \mathrm{f} \end{pmatrix} \epsilon \, SO(2,R), \qquad (3,8)$$

and taking into account the recurring feature that DM particle is actually a Bino in the major portion of the constrained MSSM parameter space

$$\mathrm{f}_{\varepsilon \rightarrow M_Z} \rightarrow 1, \; \mathrm{g}_{\varepsilon \rightarrow M_Z} \rightarrow 0, \qquad (3,9)$$

we assume for reasons of simplicity, although more complicated versions of the energy dependent solutions are possible, that the gaugino state has the following explicit form

$$|\widetilde{N}\rangle = e^{-\tau(\varepsilon - M_Z)} |\breve{B}\rangle + \sqrt{1 - e^{-2\tau(\varepsilon - M_Z)}} |\widetilde{W}^{\circ}\rangle. \qquad (3,10)$$

The scenario driven through this exponential behaviour [1] permits to monitor continuously the system at different energy stages and thus a possible physical picture of the DM state evolution. To make contact with the property (3.6), it is enough to parameterize f = $\cos\theta$ and g = $\sin\theta$ and solve to get the relation $\theta = \theta(\varepsilon)$ between the involved angle and the scale energy $\varepsilon$; later on this angle will be interpreted as the Weinberg mixing angle $\theta_w$. The parameter $\tau > 0$ is homogenous to time (inverse of energy) which in general could be used to characterize the different state configurations. When varying the scale energy and keeping it fixed, we delimit the dominating areas of each component:

| $\varepsilon$ | $\varepsilon > M_Z + \frac{Ln2}{2\tau}$ | $\varepsilon = M_Z + \frac{Ln2}{2\tau}$ | $\varepsilon < M_Z + \frac{Ln2}{2\tau}$ | $\varepsilon \sim M_Z$ |
|---|---|---|---|---|
| f, g | f $<$ g | f $=$ g | f $>$ g | f $\simeq$ 1 |
| $\widetilde{N}$ | $\Omega_{\breve{B}} < \Omega_{\widetilde{W}^{\circ}}$ | $\Omega_{\breve{B}} \sim \Omega_{\widetilde{W}^{\circ}}$ | $\Omega_{\breve{B}} > \Omega_{\widetilde{W}^{\circ}}$ | $\sim\widetilde{B}$ |

Table1: The DM compositeness evolution with the scale energy.

In this view, the DM particle lies near the electroweak scale or at least above a neighboring superparticle mass scale $M_S$ with $M_S - M_Z = \delta\varepsilon$ refers to the fine gap between the two scales.

# 4. Low energy constraints

To make this approach more predictive at weak mass scale, we use the fixed GUT relationship between the gauginos masses $M_{\breve{B}} \simeq M_{\widetilde{W}^{\circ}}/2$ as expected from renormalization group equations [23] implying $M_{\breve{B}} \geq 50GeV$. The phenomenological importance of this translation will be enhanced in what follows: the mixed state, defined as a hybridization of Bino and Wino, evolves at low energy towards the Bino, interpreted as the effective DM brick at low mass scales, with a remaining Wino content and eventually a Higgsino contribution to acount for the correct observed DM density,

$$\Omega_{\breve{B}} \simeq f|^{2}_{\varepsilon\sim M_Z},$$

$$\Omega_{\widetilde{W}^\circ} \simeq g|^{2}_{\varepsilon\sim M_Z},$$

$$\Omega_{\widetilde{H}^\circ} = \left|1 - \left(\Omega_{\breve{B}} + \Omega_{\widetilde{W}^\circ}\right)\right|. \qquad (4,1)$$

[1] The density f² could be thought of as the Boltzmann weight $e^{-\beta\varepsilon}$ with $\beta$ propotional to the inverse of temperature.

Near this scale, the resulting DM state takes the form

$$|\widetilde{N}\rangle = (1 - \tau\delta\varepsilon)|\breve{B}\rangle + \sqrt{2\tau\delta\varepsilon}|\widetilde{W}^\circ\rangle + \sqrt{\Omega_{\widetilde{H}^\circ}}. \qquad (4,2)$$

Accepting at this stage the compositeness could be treated in terms of the Weinberg mixing angle $\theta|_{\varepsilon\sim M_Z} = \theta_w$, using the parametrization : $f|^{2}_{\varepsilon\sim M_Z} = \cos\theta_w$, and $g|^{2}_{\varepsilon\sim M_Z} = \sin\theta_w$, we deduce the DM compositeness including the negligible Higgsino additional amount

$$\Omega_{\widetilde{H}^\circ} = (\tau\delta\varepsilon)^2 \qquad (4,3)$$

in such approximation when addressing DM as a pure gaugino mixture to make the analysis more economical. Straightforward calculations lead to the hybridization:

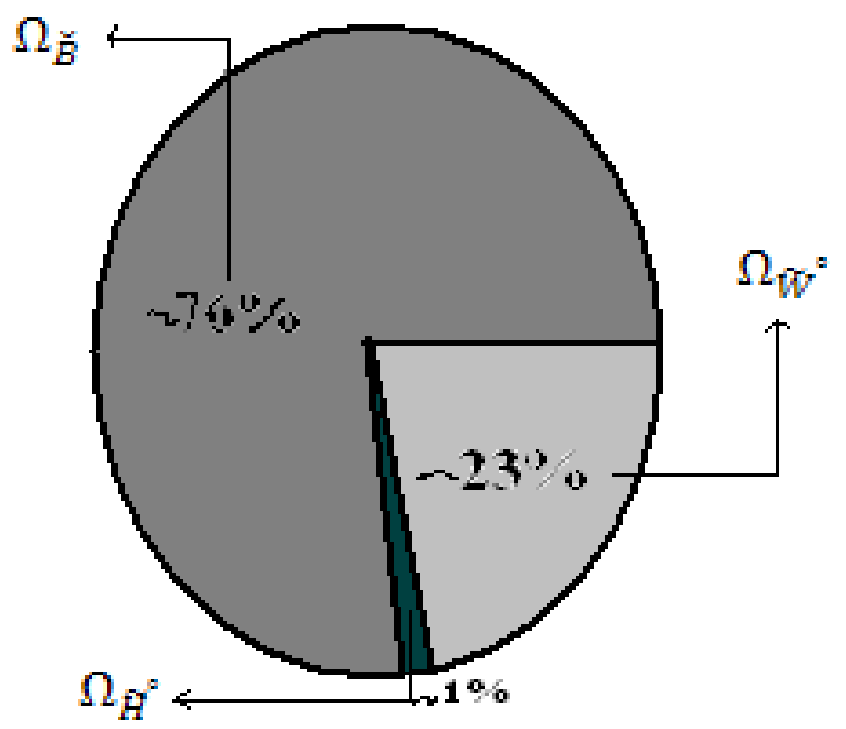

Fig1: The hybridization invloves greatly around 3/4 of Binos and 1/4 of Winos.

In this vision, ignoring the Higgsino content and using the Casimir energy momentum operator $P^2$ as well as the relationship between gaugino masses expected from renormalization group equations, we can express the mass of the DM particle (4.2) in terms of mass eigenstates as follows

$$M_{\widetilde{N}} \simeq M_{\breve{B}}\sqrt{1 + 3\sin^2\theta_w} \qquad (4,4)$$

where it depends mainly on the Bino mass. Generally, the right DM abundance (by equilibrium freeze-out) is approached by a WIMP particle lying near the electroweak scale. Indeed, such suggestion is clearly shown in this proposal where the lower bound of DM particle mass could go down till

$$M_{\widetilde{N}} \geq 65GeV. \qquad (4,5)$$

This remarkable prediction strongly suggests that such particle is tied to the electroweak scale and then should produced at the LHC. This is an interesting result for a relatively heavy elementary particle so that the previous accelerator experiments did not have enough energy to create them, whereas the Big Bang did once have energy to make them.

If Bino-like particles really make up the cold DM, with a local mass density of the order of that suspected in our neighborhood to explain the dynamics of our own galaxy, they should be distributed in a halo surrounding our galaxy with a typical speed of $v = 10^{-3}c$ [24] and would coherently scatter off nuclei in terrestrial detectors [25]. The detection of this kind of particles may be indirectly via their self-annihilation products search $\widetilde{N} + \widetilde{N} \to q\bar{q} + l\bar{l}$, or directly by studying their interaction within the detector, the tiny shocks with its atomic nuclei, with a mean energy of ~ tens of $KeV$ as recently shown by the Cryogenic Dark Matter Search experiment (CDMS II) data [26]. According to our approach, such energy is expected to be in the range

$$32{,}5\ KeV \leq E_{\widetilde{N}} < \sim MeV. \qquad (4,6)$$

It's a tiny energy deposit (recoiling energy) that is very hard to pick up against background from naturel radioactivity which is typically of $\sim MeV$. Direct search experiments seek recoil signatures of these interactions and have achieved the sensitivity to begin testing the most interesting classes of WIMP's models [27,28,29,30]. This would let place for another unspecified DM candidate. Or, the DM particle might be some particle that such Bino-like particle decays into. One possibility beyond the MSSM remains the gravitino [31,32,33]. Of course there is much to do in this path and thereby it would be important to go deeper to derive more refined results.

# 5. Concluding comments

We have seen that sensitive and economical models for supersymmetry at the TeV scale can be used as convenient templates for experimental searches. The simplest possibility is the MSSM, the popular extension of the SM fulfilling aesthetically their gaps, is recently recognized deserving to be tested experimentally. The measurements of Sparticle masses, production cross-sections, and decay modes will rule out some models for Susy breaking and lend credence to others. These measurements will be able to test the principle of R-parity conservation, the idea that Susy has something to do with the DM, and possibly make connections to other aspects of cosmology including baryogenesis and inflation. Perhaps it is not a coincidence that such particles which may solve crucial problems in particle physics also solve the DM problem. An important remark is that, from the particle physics point of view, DM may naturally be composite offering then an extra issue for interesting phenomenology.

The approach developed in this paper realizes in a simple manner the idea of compositeness in terms of gauginos wave functions hybridization. Although compositeness at high energies is somehow unlikely, nature might be kind enough to carry out small ideas such as the hybridization described in this study. Within the MSSM in the rage $M_{GUT} - M_Z$ and the analysis of section 3 an 4, the DM particle resulting from sparticles decays should be always exist abundantly to explain the large missing mass in the universe.

## Acknowledgement

We are very much grateful to referees for their valuable suggestions for the improvement of the paper.